\documentclass[sigconf,screen]{acmart}

\AtBeginDocument{%
  }

\usepackage{listings}
\usepackage{xcolor}
\usepackage{framed}
\usepackage{colortbl}
\usepackage{pifont}
\usepackage{microtype} % For better spacing and hyphenation
\usepackage{tikz}
\usetikzlibrary{arrows.meta,positioning}
\definecolor{codegreen}{rgb}{0,0.6,0}
\definecolor{codegray}{rgb}{0.5,0.5,0.5}
\definecolor{codepurple}{rgb}{0.58,0,0.82}

\lstdefinestyle{mystyle}{
    commentstyle=\color{codegreen},
    keywordstyle=\color{magenta},
    numberstyle=\tiny\color{codegray},
    stringstyle=\color{codepurple},
    basicstyle=\ttfamily\scriptsize,
    breakatwhitespace=false,         
    breaklines=true,                 
    captionpos=b,                    
    keepspaces=true,                 
    numbers=left,                    
    numbersep=5pt,                  
    showspaces=false,                
    showstringspaces=false,
    showtabs=false,                  
    tabsize=2
}

\copyrightyear{2026}
\acmYear{2026}
\setcopyright{cc}
\setcctype{by}
\acmConference[ASE '26]{Proceedings of the 41st IEEE/ACM International Conference on Automated Software Engineering}{October 12--16, 2026}{Munich, Germany}
\acmBooktitle{Proceedings of the 41st IEEE/ACM International Conference on Automated Software Engineering (ASE '26), October 12--16, 2026, Munich, Germany}
\acmDOI{10.1145/3832783.3837514}
\acmISBN{979-8-4007-2882-2/2026/10}
\acmSubmissionID{ase26main-p1862-p}
\received{2026-03-26}
\received[accepted]{2026-06-18}
\begin{document}

\title{ArkEval: Benchmarking and Evaluating Automated Code Repair for ArkTS}

\author{Bang Xie}
\orcid{0009-0004-7280-5411}
\affiliation{%
  \institution{Shanghai Jiao Tong University}
  \city{Shanghai}
  \country{China}
}
\email{xiebang-1213@sjtu.edu.cn}

\author{Senjian Zhang}
\orcid{0009-0009-5053-0994}
\affiliation{%
  \institution{Shanghai Jiao Tong University}
  \city{Shanghai}
  \country{China}
}
\email{zhangsenjian@sjtu.edu.cn}

\author{Zhiyuan Peng}
\orcid{0009-0003-6560-733X}
\affiliation{%
  \institution{Shanghai Jiao Tong University}
  \city{Shanghai}
  \country{China}
}
\email{pzy2000@sjtu.edu.cn}

\author{Wei Chen}
\orcid{0009-0006-6328-2510}
\affiliation{%
  \institution{Shanghai Jiao Tong University}
  \city{Shanghai}
  \country{China}
}
\email{chenwei8@sjtu.edu.cn}

\author{Xin Yin}
\orcid{0009-0005-3396-0571}
\affiliation{%
  \institution{Zhejiang University}
  \city{Hangzhou}
  \country{China}
}
\email{xyin@zju.edu.cn}

\author{Chenhao Ying}
\correspondingauthor
\orcid{0000-0002-2152-3826}
\affiliation{%
  \institution{Shanghai Jiao Tong University}
  \city{Shanghai}
  \country{China}
}
\email{yingchenhao@sjtu.edu.cn}

\author{Yuan Luo}
\orcid{0000-0002-3910-5286}
\affiliation{%
  \institution{Shanghai Jiao Tong University}
  \city{Shanghai}
  \country{China}
}
\email{yuanluo@sjtu.edu.cn}

\renewcommand{\shortauthors}{Xie et al.}

\begin{abstract}
  \sloppy
  Automated program repair benchmarks largely cover mature languages such as Python and Java, leaving limited support for ArkTS and OpenHarmony. We present \textbf{ArkEval}, an executable repository-level benchmark for evaluating repair on real ArkTS/OpenHarmony issue-resolution tasks. To the best of our knowledge, it is the first benchmark with this scope.

  We mined more than 6,000 pull-request and merge-request records available through July 3, 2025, from nine public repositories. Automated profiling and unanimous screening by three ArkTS/OpenHarmony developers produced 502 instances. Their reproduction tests were constructed through a six-round benchmark-wide refinement process, followed by targeted closure for the remaining six instances. We evaluated eight current models with the same localization-guided, RAG-off repair protocol. The best model compiled 54.38\% of its patches, whereas the highest behavioral Pass@1 was 19.32\%. ArkEval therefore separates build validity from behavioral repair and supplies traceable tests for studying both.
\end{abstract}

%%
%% The code below is generated by the tool at http://dl.acm.org/ccs.cfm.



\maketitle

\section{Introduction}

Large language models now support code generation and repository-level repair from natural-language descriptions \cite{brown2020language,chen2021evaluating,achiam2023gpt,xia2023keep,jimenez2024swebench}. However, influential executable benchmarks such as HumanEval, MBPP, CoderEval, DevEval, and SWE-bench remain centered on Python and/or Java \cite{chen2021evaluating,austin2021program,yu2024codereval,li2024deveval,jimenez2024swebench}. Few resources combine issue-linked revisions, ecosystem-specific builds, and executable regression tests for specialized languages.

This limitation matters for runtime-specific languages: code resembling a familiar language may still violate uncommon compiler or framework rules.

HarmonyOS application development relies on \textbf{ArkTS}, which combines strict static typing with a declarative UI model. General TypeScript code can therefore be syntactically plausible yet fail ArkTS compilation or misuse ArkUI state and lifecycle rules. Existing ArkTS benchmarks address retrieval, generation, or completion, but not executable repair of issue-linked repository revisions.

We introduce \textbf{ArkEval} to evaluate this setting. To the best of our knowledge, ArkEval is the first executable repository-level benchmark for automated repair on real ArkTS/OpenHarmony issue-resolution tasks. Because most mined revisions lacked suitable regression tests, we also developed a \textbf{Skill-guided Loop Engineering} process to construct and audit reproduction tests. The benchmark covers nine public repositories, including 149 applications represented in the official OpenHarmony sample repository. We make four contributions:
\begin{itemize}
    \item \textbf{Executable repository-level repair benchmark}: ArkEval contains 502 issue-resolution instances with traceable buggy and fixed revisions, standardized problem statements, and executable reproduction tests.
    \item \textbf{Audited test construction}: Each reproduction test passed Skill-guided generation, three independent GPT-5-series reviews, two-state execution, and final review by three human experts. Repeated audit findings were incorporated into later Skill versions.
    \item \textbf{ArkFix evaluation baseline}: The configurable baseline separates fault localization and patch generation from applicability, compilation, and behavioral testing. It also supports controlled retrieval of official ArkTS/OpenHarmony knowledge.
    \item \textbf{Eight-model evaluation}: We report results for eight current models from several providers, including a locally deployed OpenPangu model, under one evaluation protocol.
\end{itemize}

\section{Background and Motivation}

\subsection{Uneven Coverage across Programming Languages}
Code models are trained and evaluated mainly on languages with large public corpora, including Python, Java, JavaScript, and C++ \cite{zhang2024survey}. Smaller ecosystems consequently provide fewer public examples and regression tests. A second difficulty is misleading similarity: a specialized language can resemble a common one while imposing different compiler or runtime rules, causing models to transfer patterns that are plausible but invalid in the target environment.

\subsection{ArkTS as a TypeScript False Friend}
ArkTS is used for native HarmonyOS applications. It derives much of its syntax from TypeScript but restricts dynamic language features to support ahead-of-time (AOT) compilation. Two differences are especially relevant to repair.

\subsubsection{Static Typing Restrictions}
Standard TypeScript supports dynamic patterns through mechanisms such as \texttt{any} and index signatures. ArkTS restricts such patterns to maintain a predictable memory layout.
Figure \ref{fig:ts_vs_arkts} gives one example. Adding a property to an object literal is common in TypeScript, but ArkTS requires an explicit class layout. A model can therefore produce plausible TypeScript that fails the ArkTS compiler.

\begin{figure*}[h]
\centering
\begin{minipage}{0.48\textwidth}
\begin{lstlisting}[language=Java, title={(a) Valid TS (Web)}, basicstyle=\ttfamily\scriptsize, numbers=none]
// Idiomatic TS: Dynamic, flexible
let user: any = { name: "Alice" };
user.age = 30; // Added at runtime

function log(data: any) {
  console.log(data); // Runtime inspection
}
\end{lstlisting}
\end{minipage}
\hfill
\begin{minipage}{0.48\textwidth}
\begin{lstlisting}[language=Java, title={(b) Invalid in ArkTS}, basicstyle=\ttfamily\scriptsize, numbers=none]
// Compile Error: "Property 'age' does not exist"
class User {
  name: string = "";
}
let user = new User();
user.name = "Alice";
user.age = 30; // Dynamic property addition is rejected
\end{lstlisting}
\end{minipage}
\caption{Dynamic property addition through TypeScript \texttt{any} is rejected by ArkTS.}
\label{fig:ts_vs_arkts}
\Description{Side-by-side TypeScript and ArkTS examples showing that dynamic property addition through any is accepted in TypeScript but rejected by ArkTS static typing.}
\end{figure*}

\subsubsection{Declarative UI State}
ArkUI tracks component state through decorators rather than general TypeScript assignments:
\begin{itemize}
    \item \texttt{@State}: Component-internal state. Changes trigger local re-render.
    \item \texttt{@Link}: Two-way sync with a parent component. Requires precise \texttt{\$} syntax for initialization.
    \item \texttt{@Prop}: One-way sync from parent to child. Deep copy semantics.
\end{itemize}

A common error is to pass a \texttt{@State} value where a child expects either a plain \texttt{number} or a linked value. A \texttt{@Link} parameter requires the \texttt{\$this.variable} form. These data-flow rules require framework-specific reasoning beyond TypeScript syntax.

\subsection{Limited Repair Benchmarks for Emerging Languages}
Table \ref{tab:benchmark_comparison} shows that established code benchmarks focus mainly on Python and Java. SolEval extends repository-level evaluation to Solidity \cite{peng2025soleval}. Recent ArkTS benchmarks cover code retrieval, standalone generation, and repository-level completion \cite{he2026arktscodesearch,erkus2026arktsgeneration,wang2026arkrepobench}, but none evaluates executable repairs for issue-linked revisions.

\begin{table}[h]
\caption{Representative executable code benchmarks. Source repositories are reported for repository-derived tasks.}
\label{tab:benchmark_comparison}
\resizebox{\columnwidth}{!}{%
\begin{tabular}{lllrrc}
\toprule
Benchmark & Task unit & Lang. & Instances & Source repos. & Tests \\
\midrule
HumanEval \cite{chen2021evaluating} & Function synthesis & Py & 164 & -- & \ding{51} \\
MBPP \cite{austin2021program} & Function synthesis & Py & 974 & -- & \ding{51} \\
ClassEval \cite{du2023classeval} & Class generation & Py & 100 & -- & \ding{51} \\
CoderEval \cite{yu2024codereval} & Repository-context generation & Py/Java & 460 & 53 & \ding{51} \\
DevEval \cite{li2024deveval} & Repository-level generation & Py & 1,825 & 115 & \ding{51} \\
SolEval \cite{peng2025soleval} & Repository-level generation & Sol & 1,507 & 28 & \ding{51} \\
\midrule
\rowcolor{lightgray} \textbf{ArkEval (Ours)} & \textbf{Issue-linked repair} & \textbf{ArkTS} & \textbf{502} & \textbf{9} & \textbf{\ding{51}} \\
\bottomrule
\end{tabular}%
}
\end{table}

ArkTS lacks a standardized collection of issue-linked buggy revisions with executable tests. This prevents consistent model comparison, separation of localization, compilation, and behavioral failures, and reliable study of whether mainstream repair methods transfer to ArkTS.

\subsection{Scope of ArkTS Repair Evaluation}
ArkTS provides a concrete setting in which repair must account for strict typing, declarative UI state, and ecosystem-specific build tools. ArkEval is designed to measure repair within this setting. Its construction process may also be useful for other languages with sparse training data and missing regression tests, but the empirical results in this paper apply only to the evaluated ArkTS/OpenHarmony repositories.

\section{Benchmark Construction}

\begin{figure*}[t]
  \centering
  \includegraphics[width=\textwidth,page=1]{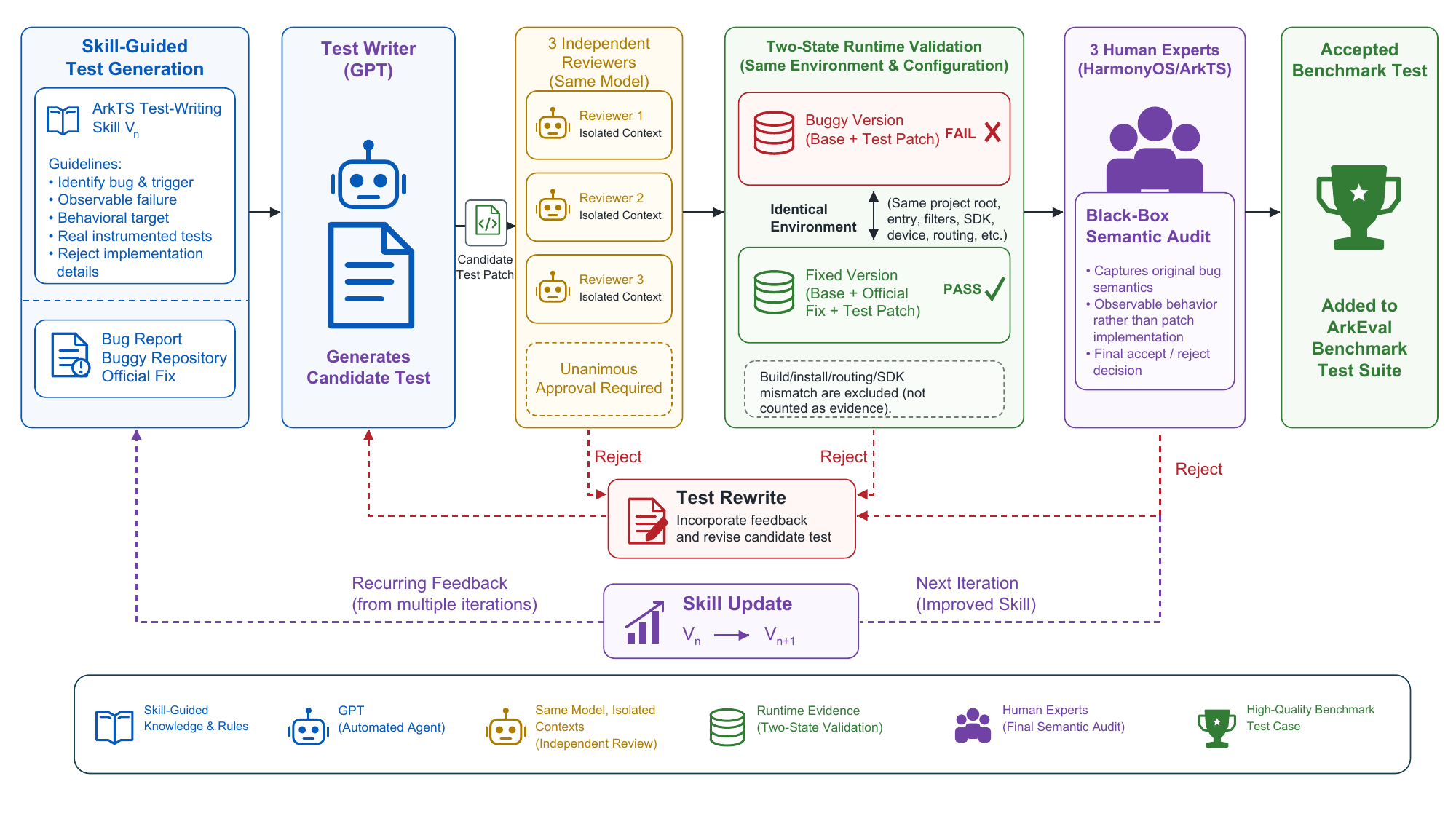}
  \caption{Loop Engineering for test generation.}
  \Description{A closed-loop workflow in which the ArkTS Test-Writing Skill guides test generation, the ArkEval Test-Validation Skill governs three independent GPT-5-series reviews and two-state runtime validation, and three human experts perform the final black-box audit. Rejections return to test rewriting, and recurring feedback updates the Skills.}
  \label{fig:benchmark_workflow}
\end{figure*}

ArkEval was constructed in five stages: repository mining, issue curation, test construction, problem standardization, and qualitative defect analysis. Figure \ref{fig:benchmark_workflow} summarizes the test-construction loop.

\subsection{Phase 1: Repository Mining and Profiling}
Given the nascent state of the ArkTS open-source community, repositories with traceable issue-resolution histories are scarce compared with mature Java or Python ecosystems. We searched GitHub and Gitee for ArkTS/OpenHarmony repositories and retained nine that satisfied four repository-level requirements: public accessibility, substantive \texttt{.ets} implementations, PR/MR records traceable to pre- and post-change revisions, and projects that could be built or exercised in an OpenHarmony environment. Mining covered more than 6,000 PR/MR records available through July 3, 2025.

The first-party OpenHarmony \texttt{applications\_app\_samples} repository was the principal source, contributing 414/502 instances (82.47\%). Unlike a single-purpose private project, it aggregates more than 400 official sample applications spanning media, device APIs, distributed behavior, and ArkUI. We resolved every modified path at its benchmark base revision to the nearest app-level Harmony project root identified by \texttt{build-profile.json5}. This produced 187 historical root paths. We then grouped internal module directories such as \texttt{entry}, \texttt{feature(s)}, \texttt{common}, and \texttt{product(s)} under their parent project and normalized legacy-to-current directory migrations. Thirty-two duplicate roots shared the same application identity across repository reorganizations, and six additional renamed roots were merged only when their functional identity was unchanged, yielding 149 distinct official sample applications. Functionally different applications remained separate. Thus, 149 is derived from the 414 first-party instances rather than from all 502 instances or from a count of modified files; the other 88 instances come from the remaining eight repositories.

The other eight repositories contributed 88 instances: \texttt{ImageKnife} (79), \texttt{md360player} (3), and \texttt{ohos\_ijkplayer}, \texttt{rdbStore}, \texttt{photo-deal-demo}, \texttt{HPRichText}, \texttt{RdbPlus}, and \texttt{Homogram} (one each). These source counts were derived directly from the 502 instance records.

\subsection{Phase 2: Constraint-Based Issue Curation}
We first profiled candidate changes automatically and then screened them semantically. We report the verifiable candidate-pool scale, both selection stages, and the final composition. Intermediate attrition counts are omitted because the original candidate list does not support them reliably.

\subsubsection{Automated Profiling and Prioritization}
A Python analysis tool first verified that a candidate exposed an obtainable code patch and traceable pre- and post-change revisions. It then profiled patch size and file composition. Rather than treating either signal as an absolute exclusion rule, we used them to prioritize localized, ArkTS-focused candidates for subsequent curation:
\begin{itemize}
    \item \textbf{Patch-size coverage}: Patch size is measured as added plus deleted lines across the developer patch, excluding diff headers. The final benchmark is dominated by localized changes: 474/502 instances (94.42\%) modify fewer than 300 lines. We retained 28 larger changes only when they had a clear issue-to-patch relationship, an identifiable behavioral target, and a reproducible executable oracle. These cases cover the long tail of repository-level repair complexity without restricting the benchmark to short patches.
    \item \textbf{ArkTS-focused prioritization}: For each candidate change, we calculated $N_{\mathrm{ets}}/N_{\mathrm{all}}$ over unique modified files, where $N_{\mathrm{ets}}$ is the number of modified \texttt{.ets} files and $N_{\mathrm{all}}$ is the total number of modified files. Only \texttt{.ets} files were classified as ArkTS files; all other modified files, including \texttt{.ts} files, documentation, logs, resources, build configurations, and generated artifacts, were treated as non-ArkTS files. We used 70\% as a prioritization threshold rather than a hard exclusion rule. In the final dataset, 417/502 instances (83.07\%) met or exceeded this threshold. The remaining 85 were retained only when the behavior-changing implementation was located in \texttt{.ets} files and the additional files provided supporting project changes.
\end{itemize}
These signals structured candidate review but did not determine inclusion on their own; the final decision was made using the semantic, scope, and reproducibility checks below.

\subsubsection{Curation and Reproducibility Check}
The same three ArkTS/OpenHarmony developers who later performed the final test-oracle audit conducted the semantic curation. Each expert reviewed candidates separately under a common rubric: the change had to be ArkTS-relevant, localized, linked to an identifiable behavior, and suitable for an executable reproduction oracle. Inclusion required unanimous approval; a substantive objection from any expert caused rejection or further review. The candidate pool was not restricted exclusively to bug-fix-labelled changes: it primarily contains bug repairs but also permits small, localized functionality enhancements.

For the 502 retained records, each expert independently recorded an initial accept-or-reject judgment before discussion. All three experts agreed on 449 records, whereas 53 contained at least one disagreement. The mean pairwise raw agreement was 93.0\%, and Fleiss' $\kappa$ was 0.79. These values describe pre-discussion judgments on the retained set, not agreement over the entire mined candidate pool. Disagreements were discussed, and retention required unanimous final approval. The experts discarded records that:
\begin{itemize}
    \item Lacked a clear semantic link between the issue description and the code change.
    \item Relied on proprietary, closed-source binary dependencies that prevented local compilation.
    \item Were duplicates or simple version bumps.
    \item Required large-scale framework refactoring, architectural migration, or broad feature redesign rather than a localized code change.
\end{itemize}
This process yielded 502 issue-resolution instances: 498 from merged PRs/MRs, two from closed records, and two from open records. Merge status was not an independent inclusion criterion. We retained the four non-merged records because their patches, version pairs, and behavioral targets were traceable and executable. For selected instances without an issue-specific regression test, we constructed a reproduction test before evaluating generated patches.

\subsection{Phase 3: Skill-Guided Test Construction}
We encoded ArkTS/OpenHarmony testing rules in two reusable Skills. The \textbf{ArkTS Test-Writing Skill} instructs the writer to identify the reported bug, its trigger, and an externally observable failure. It converts the developer fix into a behavioral target, limits edits to test paths, and requires instrumented tests for UI, Ability, lifecycle, or device behavior. It rejects assertions over source strings, ASTs, private members, exact imports, patch structure, or other fix-specific details. GPT-5.5 and GPT-5.6-Sol used this Skill to generate or revise candidate \texttt{test\_patch} files.

The \textbf{ArkEval Test-Validation Skill} defines the reviewer rubric, unanimous approval rule, two-state execution protocol, failure classes, and audit record. The three human experts, rather than the Validation Skill, made the final semantic decision.

\subsubsection{Independent GPT-5-Series Review}
Under the ArkEval Test-Validation Skill, three GPT-5-series reviewer instances assessed each candidate in isolated contexts. As the benchmark and its tests were maintained over time, later GPT-5-series checkpoints were adopted when they became available. Within each review batch, all three instances used the same checkpoint and rubric. They checked whether the test (1) triggered the reported bug, (2) exposed an observable failure, (3) distinguished the buggy and fixed versions for the reported reason, and (4) avoided the reference developer patch's private implementation. Each reviewer also explained why the base should fail and the fixed version should pass. A candidate proceeded only after all three reviewers approved it. This was not an inter-model vote.

\subsubsection{Two-State Runtime Validation}
Following the ArkEval Test-Validation Skill, unanimously approved tests were executed under the same project root, test entry, filtering conditions, and environment in two states:
\begin{center}
\texttt{base + test\_patch: FAIL} \qquad
\texttt{base + reference developer patch + test\_patch: PASS}.
\end{center}
The added test had to execute in both states; the base failure had to arise from the target bug; and the fixed-state pass had to arise from its correction. Patch-application, SDK, build, installation, routing, or test-discovery failures were recorded separately and were never counted as fail-to-pass evidence.

\subsubsection{Final Human Audit and Feedback}
The same three experts served in all six iterations. They had also curated issues in Phase 2, but in Phase 3 they entered only after agent review and runtime validation. All were ArkTS/HarmonyOS developers who had developed and tested multiple ArkTS applications.

Each expert independently checked whether the candidate was a black-box test of observable behavior and whether its simulator evidence reproduced the reported bug. Acceptance required three approvals. If one expert raised a concern, the group discussed it; an unresolved concern returned the test for revision. The experts did not write tests, act as agent reviewers, or run the validation. Rejected tests were rewritten by an agent and passed through every gate again. Repeated feedback was also added to the next Skill version.

In V1, the reviewers returned 245 of 502 candidates for revision. The remaining 257 entered runtime validation, where 99 passed the two-state check. Human experts accepted 87 of those tests and rejected 12 because they used white-box or fix-specific assertions. The 87/99 ratio measures confirmation of the combined agent-review and runtime pipeline, not the accuracy of agent review alone. It also shows that base-fail/fix-pass execution is insufficient without semantic review. For iterations V2 to V6, we report the input and final accepted count because we do not use comparable intermediate breakdowns in the analysis.

Quality control operated at both the instance and process levels. Each accepted test required unanimous agent approval, two-state execution, and unanimous expert approval. A test returned at any gate was revised and repeated all preceding gates; an earlier pass was not carried forward to the revised candidate. Recurring rejection reasons became explicit rules in the next Test-Writing and Test-Validation Skill versions.

\begin{table*}[t]
\centering
\caption{Observed six-iteration Loop Engineering results. ``Iteration rate'' uses the current iteration input as denominator.}
\label{tab:loop_engineering}
\small
\setlength{\tabcolsep}{5pt}
\begin{tabular}{lrrrrr}
\toprule
Skill/Stage & Input & Accepted & Iteration rate & Remaining & Cumulative accepted \\
\midrule
V1 & 502 & 87 & 87/502 = 17.33\% & 415 & 87 \\
V2 & 415 & 104 & 104/415 = 25.06\% & 311 & 191 \\
V3 & 311 & 109 & 109/311 = 35.05\% & 202 & 300 \\
V4 & 202 & 101 & 101/202 = 50.00\% & 101 & 401 \\
V5 & 101 & 67 & 67/101 = 66.34\% & 34 & 468 \\
V6 & 34 & 28 & 28/34 = 82.35\% & 6 & 496 \\
Post-V6 targeted closure & 6 & 6 & 6/6 = 100\% & 0 & 502 \\
\bottomrule
\end{tabular}
\end{table*}

Human feedback changed the Skill after each iteration. V2 removed white-box and fix-specific assertions. V3 required direct reproduction of the reported bug, and V4 defined project-root, suite-registration, and instrument-execution rules. V5 added rules for UI initialization, routing, waiting, and environment isolation. V6 consolidated the black-box oracle and validation handoff. The iteration acceptance rate increased from 87/502 (17.33\%) in V1 to 28/34 (82.35\%) in V6. The final six cases were revised from targeted expert feedback and passed the same review and execution gates.

\subsection{Phase 4: Issue Representation}
Public records vary in structure and detail. We preserve their title and body in consistent JSONL fields and derive a task-facing description that identifies the component, trigger, observed behavior, and behavioral target when the source record provides this evidence. We do not infer missing stack traces, reproduction steps, or localization hints. Table~\ref{tab:arkeval_instance} illustrates this representation using a retained OpenHarmony sample instance.

\begin{table}[t]
  \centering
  \caption{Example of ArkEval issue standardization.}
  \label{tab:arkeval_instance}
  \scriptsize
  \renewcommand{\arraystretch}{1.12}
  \setlength{\tabcolsep}{4pt}
  \begin{tabular}{@{}p{0.43\columnwidth}p{0.51\columnwidth}@{}}
  \toprule
  \textbf{Raw issue record} & \textbf{Standardized description} \\
  \midrule
  \textbf{Repository:} \texttt{applications\_app\_samples}\par
  \textbf{Record:} \#5960 (merged)\par
  \textbf{Title:} Pre-store PixelMap\par
  \textbf{Report:} Pre-storing PixelMap to prevent an app crash when ``Compile EXIF Message'' is selected before an image is saved.
  &
  \textbf{Component:} PixelMap sample\par
  \textbf{Trigger:} Select ``Compile EXIF Message'' before saving an image.\par
  \textbf{Observed behavior:} The application crashes.\par
  \textbf{Behavioral target:} The action must be handled without terminating the application. \\
  \bottomrule
  \end{tabular}
\end{table}

\subsection{Phase 5: Qualitative ArkTS Defect Categories}
To describe recurring ArkTS-specific failure patterns without implying an unsupported exhaustive distribution, we use three qualitative defect categories:

\begin{enumerate}
    \item \textbf{UI state desynchronization}: Imperative logic fails to trigger a declarative UI update, for example when code modifies a normal property instead of a \texttt{@State} variable. Repair requires cross-component data-flow reasoning.
    \item \textbf{Strict compile-time violations}: General TypeScript patterns, including dynamic property access, structural typing, and \texttt{any}, are rejected by the ArkTS AOT compiler.
    \item \textbf{Component lifecycle mismanagement}: Incorrect use of hooks such as \texttt{aboutToAppear} and \texttt{aboutToDisappear} can leak resources, mishandle listeners, or cause teardown failures.
\end{enumerate}

\subsection{Dataset Characteristics}
\textbf{ArkEval} contains 502 issue-resolution instances from nine repositories; the first-party OpenHarmony sample repository represents 149 applications. Tasks span reusable UI components, device integration, ArkTS typing, declarative state, lifecycles, and cross-file dependencies.

\subsection{Instance Artifacts and Provenance}
Each JSONL record uses a stable \texttt{instance\_id} and retains its organization, repository, issue/PR/MR number and text, linked issues, base revision, developer \texttt{fix\_patch}, reproduction \texttt{test\_patch}, test categories, hints, and curated defect files. The source identifiers permit reconstruction of the public record URL, while the base revision and developer patch identify the buggy and reference states. Public issue evidence, the developer change, and the independently audited reproduction test therefore remain distinct.

The blind repair input retains the issue, identifier, base revision, and model-specific localized scope, but removes the developer fix and reproduction test. Only the downstream evaluator sees the complete record to reconstruct the repository, apply model and test patches, infer the relevant HarmonyOS project root, and record application, build, and test outcomes. This split hides the reference solution and oracle while retaining score provenance.

Results are joined through \texttt{instance\_id}, not row order or path, because one repository can contain several retained revisions and renamed directories. The identifier links localization, repair trajectory, diff, optional RAG metadata, build, behavioral result, and test audit, allowing aggregate tables and failures to be traced without exposing gold information to the model.

\section{Automated Repair Evaluation Baseline}

\begin{figure*}[t]
\centering
\includegraphics[width=\textwidth]{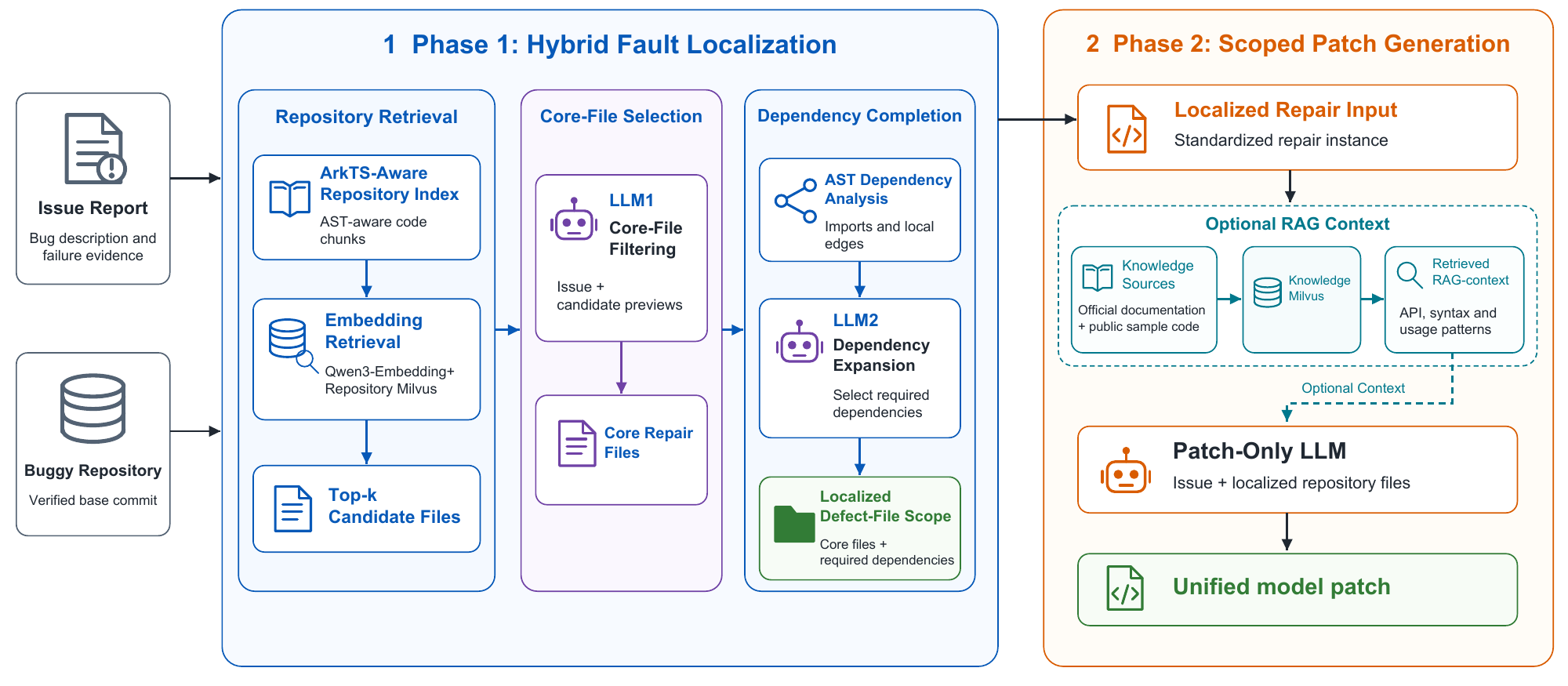}
\caption{Overview of the ArkFix localization-guided repair evaluation baseline.}
\Description{A two-phase evaluation baseline. Phase 1 retrieves candidate files by ArkTS-aware embedding, filters core files with a first model pass, and completes dependencies with static analysis and a second model pass. Phase 2 generates a unified patch from the resulting localized scope, optionally augmented with official-knowledge RAG context.}
\label{fig:repair_architecture}
\end{figure*}
 
\subsection{Task Formulation}
We formalize the Repository-Level Automated Program Repair task as follows. Given a repository $\mathcal{R}$ consisting of a set of source files $\{f_1, f_2, ..., f_n\}$ and an issue description $D$ (which may include a natural language bug report and a failure trace), the goal is to generate a patch $P$ such that the patched repository $\mathcal{R}' = \text{Apply}(P, \mathcal{R})$ satisfies two conditions:
\begin{enumerate}
    \item \textbf{Compilability}: $\text{Compile}(\mathcal{R}') \rightarrow \text{Success}$.
    \item \textbf{Correctness}: $\text{Test}(\mathcal{R}', T_{repro}) \rightarrow \text{Pass}$ and, when available, $\text{Test}(\mathcal{R}', T_{reg}) \rightarrow \text{Pass}$, where $T_{repro}$ is the reproduction test and $T_{reg}$ provides evidence that the available suite detects no regression.
\end{enumerate}
The task is repository-level because a change to one file, such as a shared UI component, can affect dependent files.

\subsection{Localization-Guided Repair Protocol}
ArkFix is a configurable evaluation baseline, not a new repair algorithm. Its localization stage follows an established engineering pattern for repository-scale semantic code search: syntax-aware chunking, embedding retrieval, and model-based candidate refinement. We adapt this pattern to ArkTS repair and evaluate it systematically rather than claiming it as a new localization algorithm.

For each instance, the repository is reset to its benchmark base revision and the indexer scans non-ignored \texttt{.ets} and \texttt{.ts} files. It extracts structural code spans, splits oversized spans at line boundaries, and limits each chunk to 2,048 characters. Qwen3-Embedding-8B encodes the chunks, which are stored in a Milvus FLAT index using L2 distance. The retrieval query is formed only from the issue title and body and any linked-issue title and body; the reference developer patch, gold defect-file annotations, and reproduction test are withheld. Milvus returns chunk-level matches, stale matches that do not correspond to the current base checkout are discarded, and each file receives the maximum similarity score of its matching chunks. Files are ranked by this score, and the reported experiments retain the Top-10 candidates.

The evaluated model then performs two constrained selection passes under shared prompts. In the first pass, it receives the issue together with candidate paths and bounded file previews and returns a non-empty JSON subset of core repair files. Static import and dependency analysis enumerates files related to this core set. In the second pass, the same model selects, again from an explicit candidate list, which dependencies must be added. The union of the core files and selected dependencies defines the localized repair scope supplied to patch generation. The embedding candidates and prompts are shared across models, whereas the two selection passes are model-specific and may therefore produce different final scopes.

The repair model receives the issue and localized files, then produces a unified diff under a shared patch-only configuration. ArkFix can retrieve official Huawei/OpenHarmony ArkTS syntax documentation, compiler-error analyses, and examples. Documentation and example-code passages are indexed separately in 512-token windows with 80-token overlap. A query combining the issue, localized files, base excerpts, and initial model observation returns at most four passages of each type and 12,000 characters. The corpus excludes reference patches, gold file annotations, reproduction tests, and generated patches. The eight main runs disable this branch for a constant information boundary and tractable $8\times502$ evaluation; RAG is evaluated separately.

Generation and evaluation remain separate in both configurations. A diff may modify only the localized scope and must apply cleanly to the benchmark base revision. We then run the same ArkTS build and reproduction test, plus the existing regression suite when available. Applicability is an artifact-validity gate, not a performance metric. Build and test results are not returned to the model within the reported Pass@1 attempt. Figure \ref{fig:repair_architecture} summarizes the protocol.

\subsection{Patch-Only Generation Contract}
Table \ref{tab:patch_contract} makes the repair model's information boundary explicit. The interface names the model-specific localized scope as \texttt{KNOWN DEFECT FILES}; this field contains predicted files, not developer-file annotations. Each instance supplies the issue, localized paths, HarmonyOS module root, current file, working directory, and ArkTS file tools. RAG-enabled runs add retrieved official-knowledge passages, while the main RAG-off runs do not. No demonstrations are inserted into either configuration.

\begin{table}[t]
\centering
\caption{Patch-only information boundary used in the main comparison.}
\label{tab:patch_contract}
\scriptsize
\setlength{\tabcolsep}{3.0pt}
\renewcommand{\arraystretch}{1.08}
\begin{tabular}{@{}p{0.25\columnwidth}p{0.69\columnwidth}@{}}
\toprule
\textbf{Boundary} & \textbf{Content} \\
\midrule
Model input & Issue, localized file paths and contents, module root, file state, and optional official-knowledge context \\
Permitted actions & Read and edit files in the localized scope; submit one unified diff \\
Withheld feedback & Reference developer patch, gold reproduction test, build, test, installation, emulator, and device outcomes \\
Downstream only & Clean checkout, patch application, \texttt{hvigor} compilation, reproduction test, and available regression suite \\
\bottomrule
\end{tabular}
\end{table}

The model is instructed to open the most relevant localized file first, inspect related localized files as needed, make the smallest functional change, and submit after the scope checker confirms a meaningful edit. Shell-based file writes, broad repository edits, test execution, builds, installation, and device actions are disabled. The interaction exposes a 100-line file window with two lines of overlap and retains the last five observations. The collected diff is filtered to the localized paths before evaluation. These constraints prevent a model from improving its single-attempt result through hidden test feedback and keep the patch-generation interface consistent across model providers.

This contract also defines failure attribution. A malformed or out-of-scope diff is an artifact failure; an applicable diff that fails \texttt{hvigor} is a compilation failure; and a compiling diff that fails the reproduction test or an available regression suite is a behavioral failure. All three statuses are retained in the instance logs, and any failed gate remains a failure under the full-denominator metrics below. Compile@1 records success only after applicability and compilation succeed, while Pass@1 additionally requires the reproduction test and, when available, the existing regression suite to pass. The separation is important because the repair model never observes any of these downstream outcomes during the reported attempt.

\subsection{ArkFix Baseline Implementation}
ArkFix implements this shared protocol by converting each model's localization output into a scoped repair instance, optionally augmenting the prompt with retrieved official knowledge, invoking the selected model under the same patch-only instructions, and recording the generated diff and evaluation outcome. The reference developer patch and gold reproduction test are withheld from the repair model in both RAG-off and RAG-on configurations. This encapsulation standardizes the information boundary and downstream gates across models. ArkFix is therefore a benchmark reference baseline and evaluation infrastructure for ArkEval, not the primary theoretical contribution of the paper; this role does not depend on whether it performs better or worse than an adapted external repair system.

ArkFix extends the SWE-agent execution substrate \cite{yang2024sweagent} with an ArkTS/OpenHarmony environment adapter; in the refreshed patch-only experiments, its build and test tools are invoked only by the downstream evaluator and their outputs are not exposed to the repair model.

\section{Evaluation}

We evaluate four questions under the same ArkFix baseline and with review-driven diagnostic pilots reported separately:

\begin{itemize}
    \item \textbf{RQ1 (fault localization accuracy)}: How accurately does ArkTS-aware embedding retrieval followed by two-pass LLM scope refinement identify repair-relevant files?
    \item \textbf{RQ2 (repair effectiveness)}: Under the same localization-guided patch-only protocol, how often do current LLMs produce compilable and behaviorally correct ArkTS patches?
    \item \textbf{RQ3 (external-knowledge RAG)}: Under a controlled repair protocol, how does official-knowledge RAG affect compilation and behavioral repair?
    \item \textbf{RQ4 (failure attribution and scope)}: What do an adapted APR baseline, developer-file localization, and cross-ecosystem pilots reveal about ArkFix's contribution and the boundaries of the ArkTS results?
\end{itemize}

\subsection{Experimental Setup}

\paragraph{Model Selection}
We selected models for provider coverage, repair capability, and stable availability in the July 2026 revision runs: \texttt{mimo-v2.5-pro}, \texttt{deepseek-v4-pro}, \texttt{kimi-k2.7-code}, \texttt{minimax-m3}, \texttt{qwen3.7-max}, \texttt{glm-5.2}, \texttt{gpt-5.6-sol}, and \texttt{openPangu-2.0-Flash}~\cite{openpangu2026flash,openpangu2026techreport}. We accessed the first seven through two OpenAI-compatible relays and deployed the last on V100/DGX hardware.

\paragraph{Main RAG-Off Comparison}
Each model's localization output defines the files supplied to repair. Patch generation uses the same no-demonstration, patch-only instructions, temperature 0, top-$p$ 1.0, at most 50 agent steps, and up to three infrastructure retries. It cannot call compilation, tests, devices, or external-documentation RAG. We disabled RAG because retrieval and longer prompts increase per-instance latency, and because one information boundary makes the $8\times502$ comparison easier to interpret. Every patch then passes through the same applicability, compilation, and behavioral-test gates.

\paragraph{Separate RAG-Enabled Comparison}
We evaluate official-knowledge RAG on \texttt{gpt-5.6-sol}, \texttt{deepseek-v4-pro}, \texttt{minimax-m3}, and \texttt{MiniMax M2.5} under the same gates. The corpus emphasizes official ArkTS constraints that differ from TypeScript. \texttt{MiniMax M2.5}, used in the earlier RAG ablation, is distinct from \texttt{mimo-v2.5-pro} and \texttt{minimax-m3}; its result is not merged with either refreshed-model row.

\paragraph{Review-Driven Diagnostic Pilots}
Three earlier pilots address questions that the main eight-model comparison cannot isolate. A cross-ecosystem pilot applies \texttt{MiniMax M2.5} to 50 TypeScript and 50 Android Kotlin/Java repair instances. An adapted ThinkRepair comparison \cite{yin2024thinkrepair} provides an external APR reference point. A 24-instance localization pilot replaces ArkFix's predicted files with the files modified by the developer patch. These pilots use different subsets or configurations and are therefore reported as diagnostics rather than merged with the main leaderboard.

The Android pilot was constructed from 23 repositories: automatic mining produced 1,819 candidates, reproducibility screening retained 200, and 50 were evaluated in the reported run. The TypeScript pilot likewise reports a fixed 50-instance set. Their ecosystem-specific retrieval sources and build environments differ, so they are scope diagnostics rather than controlled language comparisons. The ThinkRepair rows use ArkEval's compilation and behavioral criteria but are adaptations, not bit-for-bit reproductions. In the 24-instance localization pilot, repair and evaluation were fixed and only the supplied scope was replaced by developer-modified files.

\paragraph{Patch Evaluation}
We evaluated each patch in four steps:
\begin{enumerate}
    \item \textbf{Environment configuration}: We inspected each repository's \texttt{build-profile.json5} and README to select its SDK and dependencies.
    \item \textbf{Patch application}: A script checked out the issue revision and applied the generated diff.
    \item \textbf{Compilation}: The \texttt{hvigor} build rejected patches with syntax, type, or project-configuration errors.
    \item \textbf{Behavioral testing}: Compiling patches were scored automatically by the fixed executable reproduction test and, when available, the existing regression suite. Expert inspection of simulator evidence was confined to benchmark test construction and never used to score generated patches.
\end{enumerate}

\subsection{Metrics}
We report localization, compilation, and behavioral repair metrics:
\begin{itemize}
    \item \textbf{Fault localization metrics}: For each issue, we compare the ranked predicted file list with the curated repair-relevant file set derived from the reference developer patch. We report Exact Match (EM), Hit@1, Hit@5, Complete@5, mean reciprocal rank (MRR), mean average precision (MAP), per-issue macro-averaged set F1 (Macro-F1), and the average number of unique predicted files (Avg. Files). EM requires equality of the predicted and gold sets; Hit@$k$ requires at least one gold file in the first $k$ predictions; and Complete@5 requires all gold files to occur in the first five predictions. MRR scores the rank of the first relevant file, whereas MAP accounts for the ranks of all gold files.
    \item \textbf{Compilation rate (Compile@1)}: The percentage of generated patches that complete the configured \texttt{hvigor} build successfully.
    \item \textbf{Repair success rate (Pass@1)}: The percentage of patches that compile and pass the executable reproduction test and, when present, the existing regression suite. Most projects provide no such suite, so regression assurance varies; all outcomes remain automatically scored.
\end{itemize}

All rates use the full stated evaluation set as denominator unless a table explicitly identifies a pilot subset. Compile@1 does not condition on successful patch generation, and Pass@1 does not condition on compilation; an unsuccessful earlier gate therefore remains a failure in the end-to-end denominator. This convention prevents a model with few applicable or compilable outputs from appearing stronger through selective conditioning. Conversely, patch applicability is reported in the instance logs as an artifact-validity status rather than as a headline metric because a valid submitted artifact is a prerequisite for evaluation. Localization percentages use one vote per issue, so repositories with more files do not receive greater weight, while Avg. Files is reported alongside accuracy metrics to expose gains obtained by returning broader scopes.

The three pilot tables preserve their own denominators: 50 TypeScript instances, 50 Android instances, and 24 localization-intervention instances. Their percentages are not pooled with the 502-instance leaderboard. RAG deltas are percentage-point differences between paired configurations on the same 502 instances, computed from the integer counts before the displayed percentages are rounded; they are not relative percentage improvements. These denominator and pairing rules are applied consistently in the result tables and released aggregation scripts.

\subsection{Reproducibility Controls}
Each model run is represented by a batch configuration and per-instance output rather than by a manually assembled score table. The batch metadata fixes the dataset, exposed model identifier, prompt configuration, temperature, top-$p$, step limit, retry limit, RAG mode, and retrieval limits. It also records the model-specific localized scope associated with each \texttt{instance\_id}. The released aggregation scripts join these identifiers to the 502 benchmark records and reject missing, duplicate, or extra identifiers before computing a result. Thus, each headline percentage uses the same declared set of 502 tasks even though the models can select different final repair scopes.

Evaluation reconstructs a clean checkout at the recorded base revision for every instance. The evaluator applies only the submitted unified diff and benchmark test patch, infers the relevant HarmonyOS project from repository build files, and executes the configured \texttt{hvigor} task before behavioral tests. Patch application, project discovery, dependency preparation, compilation, test discovery, reproduction-test execution, and available-regression execution are stored as distinct statuses. This prevents an unavailable SDK, an unapplied patch, or an undiscovered test from being counted as a behavioral repair.

The model receives one patch-generation attempt: downstream build and test outcomes are never returned to it. Infrastructure retries rerun the same configured task after a worker or transport failure; they do not provide semantic feedback or permit a revised patch. For RAG-enabled runs, the replication package preserves the documentation corpus, chunk sidecar, index settings, retrieved passage metadata, and injected context. For relay-hosted models, we report the exact identifier exposed by the evaluation service and do not infer an undisclosed provider-side checkpoint beyond that identifier. These controls make the reported information boundary and aggregation procedure inspectable while keeping provider-internal implementation details outside the paper's claims.

\subsection{Results}

\subsubsection{Fault Localization Accuracy (RQ1)}
We evaluated all eight localization result files on the same 502 issues. Before model-based refinement, the shared embedding stage retrieved ten candidate files per issue and placed at least one gold repair-relevant file in this set for 304/502 issues (60.56\%). This is candidate-stage Hit@10 rather than final localization accuracy. Each model-specific result file contains 502 unique instance identifiers with no missing, extra, or empty predictions, and its two recorded localization fields agree row by row. Table \ref{tab:localization_results} combines strict set agreement, ranked retrieval, coverage, and output-size measures. All metrics except Avg. Files are reported as percentages; each issue receives equal weight. The gold set denotes curated repair-relevant files from the reference developer patch rather than every file modified in that patch.

\begin{table*}[t]
\centering
\caption{File-level localization performance on 502 ArkEval issues. Higher is better except for Avg. Files, which reports prediction-set size rather than quality.}
\label{tab:localization_results}
\footnotesize
\setlength{\tabcolsep}{2.6pt}
\renewcommand{\arraystretch}{1.08}
\begin{tabular*}{\textwidth}{@{\extracolsep{\fill}}lrrrrrrrr}
\toprule
\textbf{Model} & \textbf{EM} & \textbf{Hit@1} & \textbf{Hit@5} & \textbf{Complete@5} & \textbf{MRR} & \textbf{MAP} & \textbf{Macro-F1} & \textbf{Avg. Files} \\
\midrule
mimo-v2.5-pro & 7.17 & 40.04 & 53.78 & \textbf{24.50} & 46.17 & 30.61 & 29.14 & 4.16 \\
deepseek-v4-pro & 11.16 & 39.44 & 52.99 & 21.91 & 45.11 & 29.30 & 30.53 & 4.02 \\
kimi-k2.7-code & 10.76 & 44.42 & \textbf{55.38} & 23.11 & \textbf{49.23} & \textbf{31.97} & 32.77 & 2.36 \\
minimax-m3 & 4.98 & 35.26 & 53.59 & 23.71 & 43.16 & 29.38 & 25.14 & 6.42 \\
qwen3.7-max & 13.55 & 44.42 & 54.18 & 22.91 & 49.00 & 31.42 & 33.24 & 2.85 \\
glm-5.2 & 12.75 & 41.83 & 51.59 & 22.51 & 46.44 & 30.27 & 31.75 & 3.73 \\
gpt-5.6-sol & \textbf{15.54} & \textbf{45.22} & 53.19 & 22.11 & 48.96 & 31.50 & \textbf{34.87} & 1.63 \\
openPangu-2.0-Flash & 7.77 & 40.04 & 46.02 & 18.33 & 42.61 & 25.74 & 24.96 & 2.47 \\
\bottomrule
\end{tabular*}
\end{table*}

No model leads every localization metric. \texttt{gpt-5.6-sol} has the highest EM (15.54\%), Hit@1 (45.22\%), and Macro-F1 (34.87\%), with 1.63 predicted files on average. \texttt{kimi-k2.7-code} leads Hit@5 (55.38\%), MRR (49.23\%), and MAP (31.97\%). \texttt{mimo-v2.5-pro} leads Complete@5 (24.50\%). List size affects these comparisons: \texttt{minimax-m3} reaches 53.59\% Hit@5 but only 4.98\% EM while predicting 6.42 files on average. We therefore report both ranked and set-level metrics.

Five properties of the retrieval protocol may limit localization:

\begin{enumerate}
    \item \textbf{Chunk representation}: ArkTS-aware syntax chunks preserve more local structure than whole-file truncation, but a maximum chunk length of 2,048 characters can still split long-range dependencies and dilute file-level score aggregation.
    \item \textbf{Embedding semantic gap}: Vector similarity may fail to capture the semantic link between brief issue descriptions and the corresponding code implementation, reducing retrieval quality.
    \item \textbf{Top-$k$ search limit}: Retrieval was limited to the Top-10 files in the reported localization runs, so correct files ranked lower were excluded. A larger search scope might improve recall but would increase downstream context and inference cost.
    \item \textbf{LLM filtering inaccuracy}: The two LLM passes can incorrectly remove a relevant core file or decline a necessary dependency. Static import analysis narrows dependency candidates but cannot recover files omitted before dependency expansion.
    \item \textbf{Issue description quality}: Many first-party records had brief descriptions. Prompt-based expansion cannot recover details absent from the source record.
\end{enumerate}

\subsubsection{Repair Effectiveness (RQ2)}
Table \ref{tab:main_results} reports the completed results for eight models on all 502 instances. Each cell gives the number of successful instances and the corresponding percentage of the full benchmark. Patch applicability is enforced before scoring and is not treated as a performance metric.

\begin{table}[t]
\centering
\caption{Repair results on 502 ArkEval instances, reported as $n$ (\%).}
\label{tab:main_results}
\footnotesize
\setlength{\tabcolsep}{3.5pt}
\renewcommand{\arraystretch}{0.96}
\begin{tabular*}{\columnwidth}{@{\extracolsep{\fill}}lrr}
\toprule
\textbf{Model} & \textbf{Compile@1, $n$ (\%)} & \textbf{Pass@1, $n$ (\%)} \\
\midrule
\texttt{mimo-v2.5-pro} & 268 (53.39) & 30 (5.98) \\
\texttt{deepseek-v4-pro} & 208 (41.43) & 26 (5.18) \\
\texttt{kimi-k2.7-code} & 62 (12.35) & 16 (3.19) \\
\texttt{minimax-m3} & 223 (44.42) & 36 (7.17) \\
\texttt{qwen3.7-max} & 187 (37.25) & 31 (6.18) \\
\texttt{glm-5.2} & \textbf{273 (54.38)} & 42 (8.37) \\
\texttt{gpt-5.6-sol} & 172 (34.26) & \textbf{97 (19.32)} \\
\texttt{openPangu-2.0-Flash} & 170 (33.86) & 20 (3.98) \\
\bottomrule
\end{tabular*}
\end{table}

Compilation and behavioral repair produce different rankings. \texttt{glm-5.2} has the highest Compile@1 at 273/502 (54.38\%), followed by \texttt{mimo-v2.5-pro} at 268/502 (53.39\%). \texttt{gpt-5.6-sol} compiles 172/502 patches (34.26\%) but passes the behavioral gate on 97/502 instances (19.32\%). The next-highest Pass@1 is 42/502 (8.37\%) for \texttt{glm-5.2}. Across models, Pass@1 ranges from 3.19\% to 19.32\%. Compilation alone is therefore insufficient as a proxy for behavioral repair on ArkEval.

\subsubsection{External-Knowledge RAG (RQ3)}
Table \ref{tab:rag_results} reports paired RAG-enabled and RAG-off comparisons for four models. The RAG-enabled runs use the same localization output and downstream compilation and behavioral gates as their matched controls; only the official-knowledge context is added during patch generation.

\begin{table}[t]
\centering
\caption{Official-knowledge RAG results and changes from matched RAG-off runs.}
\label{tab:rag_results}
\footnotesize
\setlength{\tabcolsep}{1.2pt}
\renewcommand{\arraystretch}{0.96}
\begin{tabular*}{\columnwidth}{@{\extracolsep{\fill}}lrrrr}
\toprule
\textbf{Model} & \textbf{Compile@1} & \textbf{$\Delta_C$} & \textbf{Pass@1} & \textbf{$\Delta_P$} \\
\midrule
\texttt{gpt-5.6-sol} & 231 (46.02) & +11.75 & 105 (20.92) & +1.59 \\
\texttt{deepseek-v4-pro} & 251 (50.00) & +8.57 & 41 (8.17) & +2.99 \\
\texttt{minimax-m3} & 260 (51.79) & +7.37 & 52 (10.36) & +3.19 \\
\texttt{MiniMax M2.5} & 145 (28.88) & +4.38 & 14 (2.79) & +0.80 \\
\bottomrule
\end{tabular*}
\end{table}

The observed Compile@1 increases are 4.38--11.75 percentage points, and the observed Pass@1 increases are 0.80--3.19 points. \texttt{gpt-5.6-sol} records the largest compilation increase, from 172/502 (34.26\%) to 231/502 (46.02\%), whereas \texttt{minimax-m3} records the largest behavioral increase, from 36/502 (7.17\%) to 52/502 (10.36\%). All four paired comparisons move in the same direction, although the comparison remains descriptive and limited to four models and one official-knowledge corpus. Table values are RAG-enabled $n$ (\%); $\Delta_C$ and $\Delta_P$ denote percentage-point changes from the matched RAG-off runs.

\subsubsection{Review-Driven Diagnostic Experiments (RQ4)}

\paragraph{Cross-Ecosystem Scope Check}
Table \ref{tab:cross_ecosystem_pilot} places the ArkTS result beside two small repair pilots conducted with the same \texttt{MiniMax M2.5} model. The TypeScript pilot achieves 34/50 Compile@1 and 12/50 Pass@1, whereas the Android Kotlin/Java pilot achieves 21/50 and 3/50. The corresponding ArkTS/OpenHarmony run reaches 145/502 and 14/502. The Android result is reported after the 23-repository candidate and reproducibility screening described above, not as a sample drawn from ArkEval. These datasets differ in repositories, issue distributions, build systems, and task composition, so the table is not a controlled language ranking and does not support direct transfer of results across ecosystems.

\begin{table}[t]
\centering
\caption{Cross-ecosystem repair pilots with MiniMax M2.5.}
\label{tab:cross_ecosystem_pilot}
\footnotesize
\setlength{\tabcolsep}{2.7pt}
\begin{tabular*}{\columnwidth}{@{\extracolsep{\fill}}lrrr}
\toprule
\textbf{Ecosystem} & \textbf{$N$} & \textbf{Compile@1} & \textbf{Pass@1} \\
\midrule
TypeScript & 50 & 34 (68.00) & 12 (24.00) \\
Android Kotlin/Java & 50 & 21 (42.00) & 3 (6.00) \\
ArkTS/OpenHarmony & 502 & 145 (28.88) & 14 (2.79) \\
\bottomrule
\end{tabular*}
\end{table}

\paragraph{Adapted APR Baseline}
Table \ref{tab:thinkrepair_pilot} reports an archived, non-paired diagnostic alongside an adapted ThinkRepair configuration \cite{yin2024thinkrepair}. The archived summary preserves percentages but not the integer counts, paired instance mapping, or common denominator needed for a direct comparison. We therefore report these values only as historical context and draw no superiority claim from their differences.

\begin{table}[t]
\centering
\caption{Archived non-paired adapted APR diagnostic (\%).}
\label{tab:thinkrepair_pilot}
\footnotesize
\setlength{\tabcolsep}{2.7pt}
\begin{tabular*}{\columnwidth}{@{\extracolsep{\fill}}llrr}
\toprule
\textbf{Workflow} & \textbf{Model} & \textbf{Compile@1} & \textbf{Pass@1} \\
\midrule
ArkFix + RAG & MiniMax M2.5 & 28.88 & 2.79 \\
ThinkRepair-adapted & MiniMax M2.5 & 28.27 & 1.99 \\
ThinkRepair-adapted & Claude Sonnet 4.5 & 29.08 & 2.59 \\
\bottomrule
\end{tabular*}
\end{table}

\paragraph{Developer-File Localization}
To separate localization errors from patch-generation errors, we reran 24 instances after replacing ArkFix's predicted scope with the files modified by the developer patch. Pass@1 increased from 1/24 (4.2\%) to 11/24 (45.8\%). The intervention yielded ten additional successful repairs on this subset. However, 13/24 instances still failed even with developer-file scope, so correct files do not eliminate ArkTS compilation and semantic-repair difficulty.

\begin{table}[t]
\centering
\caption{Developer-file localization pilot.}
\label{tab:developer_file_pilot}
\footnotesize
\begin{tabular*}{\columnwidth}{@{\extracolsep{\fill}}lrr}
\toprule
\textbf{Repair scope} & \textbf{$N$} & \textbf{Pass@1} \\
\midrule
ArkFix-predicted files & 24 & 1 (4.2\%) \\
Developer-modified files & 24 & 11 (45.8\%) \\
\bottomrule
\end{tabular*}
\end{table}

\section{Discussion}

\subsection{What the Results Show for ArkTS Repair}
Current models solve relatively few ArkTS/OpenHarmony tasks under strict AOT compilation and behavioral testing. None of the eight models exceeds 20\% Pass@1, although the best Compile@1 exceeds 54\%. For \texttt{gpt-5.6-sol}, Compile@1 exceeds Pass@1 by 14.94 percentage points. The models that lead the two metrics also differ, so compilation cannot substitute for behavioral validation.

The results support three evaluation decisions:
\begin{enumerate}
    \item \textbf{Test construction requires semantic audit}: Repeated human findings were converted into test-writing rules. The iteration acceptance rate changed from 87/502 in V1 to 28/34 in V6. The 12 first-round rejections also show that runtime evidence alone does not establish a valid black-box oracle.
    \item \textbf{Localization and repair require separate metrics}: Retrieving a relevant file does not guarantee a compilable or behaviorally correct patch. Separate localization, compilation, and test metrics identify the stage at which a run fails.
    \item \textbf{Official-knowledge RAG suggests a role for domain-specific context}: All four paired comparisons show observed increases, with larger changes in compilation than behavioral success. The official ArkTS constraints in the corpus suggest that some failures reflect insufficient language-specific context, but error-level classification is needed to attribute changes to syntax violations.
\end{enumerate}

\subsection{Attributing Failures across the Pipeline}
The combined results locate several distinct bottlenecks rather than supporting a single explanation for low Pass@1. Before model refinement, embedding retrieval includes at least one gold repair-relevant file in the Top-10 set for 304/502 issues (60.56%). Candidate recall is therefore substantial but incomplete, and the subsequent two model passes can also remove a relevant file or add distracting dependencies. The model-specific localization metrics in Table \ref{tab:localization_results} should consequently be read as properties of the complete retrieval-and-refinement pipeline, not of Qwen3-Embedding-8B alone.

The 24-instance scope intervention provides a second separation. Replacing predicted files with developer-modified files increases Pass@1 from 1/24 to 11/24, yielding ten additional successful repairs on this subset. However, 13 instances still fail under developer-file scope. Those remaining outcomes can arise from ArkTS compilation constraints, insufficient issue information, incorrect patch logic, regression failures, or the fact that developer-modified files are an informative but not necessarily minimal semantic repair scope. The pilot therefore supports localization as a major bottleneck without reducing the end-to-end problem to localization alone.

Finally, compilation and behavioral success respond differently to domain context. Official-knowledge RAG produces observed Compile@1 increases of 4.38--11.75 percentage points and Pass@1 increases of 0.80--3.19 points. The larger compilation changes are consistent with retrieved ArkTS rules helping models avoid language and API violations. They do not prove that every prevented failure is syntactic, because retrieved context may also change localization interpretation or patch logic. Taken together, the localization intervention, RAG comparison, and compile-to-pass gap motivate reporting each gate separately and retaining bounded claims about model capability.

\subsection{Threats to Validity}
The study has six main limitations:
\begin{itemize}
    \item \textbf{Correlated agent errors}: Within each review batch, the three reviewers are isolated instances of the same GPT-5-series checkpoint, so their errors may be correlated. Updating the checkpoint during later maintenance does not remove this within-batch dependence. The 87/99 human-confirmation ratio describes the combined agent-review and runtime pipeline, not agent review alone.
    \item \textbf{Two-state semantics}: Base-fail/fix-pass execution is necessary but insufficient for semantic validity. In V1, experts rejected 12 runtime-passing tests that inspected white-box or fix-specific details. We therefore retain final human review.
    \item \textbf{Runtime coverage and environment}: The first-round runtime report covered 257 of 502 candidates. The 158 failed or blocked outcomes include test and environment failures, which we record separately. Most projects lack existing regression suites, yielding heterogeneous regression assurance despite automatic scoring.
    \item \textbf{Localization constraints}: The index uses ArkTS-aware chunks of at most 2,048 characters and retains the Top-10 files before LLM refinement. Chunk boundaries, score aggregation, and this candidate limit may reduce recall.
    \item \textbf{RAG evaluation boundary}: All eight main runs disable RAG; the paired comparison covers four models and one official-knowledge corpus. Its observed changes may depend on corpus coverage, query construction, retrieval relevance, and model use of context. Without an error-level taxonomy, we cannot isolate avoided ArkTS syntax violations from other contextual effects.
    \item \textbf{Dataset scope}: ArkEval covers nine repositories and 149 applications, but 414 of 502 instances come from the first-party OpenHarmony sample repository. These samples are authoritative and executable, yet they may not reflect the scale, dependencies, or maintenance practices of industrial applications. Moreover, 474/502 patches change fewer than 300 lines, and framework-scale refactorings were excluded. The results therefore characterize localized issue resolution rather than architectural repair. They apply to the evaluated open-source ArkTS/OpenHarmony projects and do not establish transfer to other languages or mobile ecosystems.
\end{itemize}

\section{Related Work}

\subsection{Benchmarks for Automated Program Repair}
APR benchmarks pair defective programs with executable tests. Defects4J supports Java repair systems such as GenProg and SimFix \cite{just2014defects4j,weimer2009automatically,le2012genprog,jiang2018shaping}; QuixBugs and HumanEval cover smaller algorithmic or synthesis tasks \cite{lin2017quixbugs,chen2021evaluating,NEURIPS2021_ea96efc0,prenner2021automatic}. SWE-bench targets repository-level GitHub issues \cite{jimenez2024swebench}, and SolEval extends executable evaluation to Solidity \cite{peng2025soleval}. None covers issue-linked ArkTS/OpenHarmony repair.

\subsection{LLM-Driven Code Repair}
CoCoNut, CURE, and DLFix formulate repair as neural translation \cite{lutellier2020coconut,jiang2021cure,li2020dlfix,tufano2019empirical,white2019sorting}. Later systems add conversational models, retrieval, execution, or textual feedback \cite{chen2021evaluating,achiam2023gpt,xia2023keep,fan2023automated,nashid2023retrieval,parvez2021retrieval,olausson2023demystifying}, but transfer to specialized ecosystems remains uncertain \cite{austin2021program,zhang2024survey}.

\subsection{Software Engineering for Mobile Ecosystems}
Prior work studies Android compatibility repair and GUI-agent environments \cite{li2022automating,zhang2024mobile}. Existing ArkTS benchmarks evaluate retrieval, generation, or repository completion \cite{he2026arktscodesearch,erkus2026arktsgeneration,wang2026arkrepobench}, while HapRepair evaluates an OpenHarmony method on a dedicated test set \cite{lin2025haprepair}. None provides real issue-linked revisions paired with executable reproduction tests.

Retrieval supplies documentation or code context to generation \cite{lewis2020retrieval,guu2020retrieval,borgeaud2022improving,zhou2023docprompting}; ArkFix evaluates official-knowledge retrieval separately from its main RAG-off comparison.

\section{Conclusion}

We presented \textbf{ArkEval}, to the best of our knowledge the first executable repository-level benchmark for real ArkTS/OpenHarmony issue-resolution tasks, with 502 traceable version pairs and reproduction tests built through Skill-guided generation, two-state execution, and human audit. Under the shared RAG-off baseline, \texttt{glm-5.2} leads Compile@1 at 54.38\%, whereas \texttt{gpt-5.6-sol} leads Pass@1 at 19.32\%, showing that compilation is an insufficient proxy for repair. Official-knowledge RAG improves both metrics in the four-model comparison. Results remain limited to nine repositories and the patch-only protocol; future work can study verified ArkTS data, DevEco Studio integration, and learning-based repair.

\begin{acks}
  This work was supported in part by the National Natural Science Foundation of China (NSFC) under Grant 62402313. We thank the reviewers for their comments.
\end{acks}

\section*{Data Availability Statement}
The ArkEval replication package is available on Figshare \cite{replication_package}.

\bibliographystyle{ACM-Reference-Format}
\bibliography{references_v75}

\end{document}